# Crystal structures and the sign reversal Hall resistivity in iron-based superconductors $Li_x(C_3H_{10}N_2)_{0.32}FeSe$ (0.15<x<0.4)

Ruijin Sun[1,2], Shifeng Jin[1,3*], Jun Den[1,2], Munan Hao[1,2], Linlin Zhao[1,2], Xiao Fan[1,2], Xiaoning Sun[1,2], Jiangang Guo[1,2], Lin Gu[1,2*], Xiaolong Chen[1,3,4*]

[1] *Beijing National Laboratory for Condensed Matter Physics, Institute of Physics, Chinese Academy of Sciences, Beijing 100190, China.*

[2]*University of Chinese Academy of Sciences, Beijing 100049, China.*

[3] *School of Physical Sciences, University of Chinese Academy of Sciences, Beijing 101408, China.*

[4] *Collaborative Innovation Center of Quantum Matter, Beijing, 100190, China.*

**Abstract**

We report the crystal structures, superconductivity and normal state properties of two iron-based materials, $Li_{0.15}(C_3H_{10}N_2)_{0.32}FeSe$ (*P*-4) and $Li_x(C_3H_{10}N_2)_{0.32}FeSe$ (*P*4/nmm, 0.25<x<0.4) with superconducting transition temperatures from 40 ~ 46 K. The determined crystal structures revealed a coupling between Li concentration and the orientation of 1,3-Diaminopropane molecules within the hyper expanded FeSe layers. Further fitting on resistivity in terms of the Lawrence-Doniach model suggests the two superconductors belong to the quasi-two dimensional (2D) system. With increasing temperatures, a universal sign reverse of Hall coefficient ($R_H$) from negative to positive is observed at ~ 185 K in both superconducting phases, regardless of their differences in crystal structures and doping levels. First principle calculations revealed the increase in FeSe layer distance will reconstruct the Fermi surface and generate a new hole pocket around Γ point in the Brillouin Zone. Our findings support that the increase in two dimensionalities will leads to a temperature induced Lifshitz transition in electron doped FeSe superconductors.

## I. Introduction

The Fermi surface (FS) topology have been considered as a key ingredient in understanding the mechanism of the iron-based superconductors (FeSCs). As revealed by angle-resolved photoemission spectroscopy (ARPES) and band structure calculations, the superconductivity and normal state of FeSCs are governed by their electronic structure involving the Fe 3d orbitals crossing the Fermi energy. In iron-pnictide materials, superconductivity generally emerges close to the disappearance of an antiferromagnetic (AFM) ordered state [1], and the maximum Tc is achieved with the presence of both hole and electron like pockets near the Brillouin zone center (Γ point) and corners (M point), respectively [2]. It is hence proposed that the electron scattering between the hole and electron pockets favored by the antiferromagnetic fluctuations is responsible for electron pairing in the iron-based superconductors [3, 4]. This picture, however, is recently challenged with the discovery of several electrons doped FeSe-derived systems, including $K_xFe_{2-y}Se_2$ [5, 6], (Li,Fe)OHFeSe [7, 8, 9], molecular intercalated FeSe [10, 11, 12, 13, 14] as well as monolayer FeSe [15, 16], wherein only the 'electron pockets' are observed near the fermi surface. The findings raise a fundamental question of whether iron pnictides and chalcogenides have different pairing mechanisms, and hence have attracted intensive research interests. Then, a coming question is whether it is possible to bring the missing hole pockets back into the electron-doped FeSe-derived superconductors, subsequent investigations may help bridge up the essential physics between the two large family of FeSCs.

Without electron doping, FeSe is a typical compensated semimetal with its FS consisting of both hole and electron bands, in analogue to iron-pnictide materials [2,17]. In fact, the hall coefficient of bulk FeSe changed sign several times with elevating temperature [18]. A notable feature of this compound is it undergoes a 'nematic' structural transition at Ts ∼ 87 K but does not order magnetically at any temperature [17]. However, high pressure can effectively induce the missing static magnetic order in undoped FeSe, and significantly increased the Tc up to 37 K [19, 20]. Recently, Sun et al. reported the normal-state Hall resistivity of FeSe also changed sign from negative to positive under high pressure [21], demonstrating enlarged hole pockets and enhanced interband spin fluctuations, in analogue to the observations in high-Tc pnictide systems. In sharp contrast, the sufficiently high electron doping ($>10^{21}/cm^3$) in intercalated FeSe system give rise to a FS consist of only 'electron pocket', distinct to pnictides as well as undoped FeSe [6, 8, 16]. Moreover, subsequent Hall resistivity measurements on $A_xFe_{2-y}Se_2$ and (Li,Fe)OHFeSe[22, 23, 24, 25] indicates neither the elevating temperature nor applied extreme high pressure will leads to sign reversal of Hall coefficient, indicating it is nontrivial to induce a reverse Lifshitz transition and bring the missing 'hole pockets' back into the intercalated FeSe systems. On the other hand, Y. Sun et al. demonstrated a sign reversal of Hall resistance with elevating temperature in few layer FeSe films deposited on $SiTO_3$ substrate[26]. More intriguing, the sign reversal temperature is found directly correlated to the layer thickness, suggesting the Lifshitz transition could be related to the reduced dimensionality in the FeSe system. The findings strongly motivate further investigations on the interplays between the dimensionality of the crystal structure, temperature, carrier doping and the Fermi surface topology in hyper-expanded FeSe systems, for example, the large organic molecular intercalated FeSe superconductors.

In this study, we report the crystal structures, superconductivity and Hall resistivity of 1,3-Diaminopropane molecular intercalated FeSCs, $Li_x(C_3H_{10}N_2)_{0.32}FeSe$ (x = 0.15, 0.2~0.4), the orbital resolved electronic structure of this system is also obtained based on experimental structural values at low temperature. Our results identified two new high-$T_c$ superconducting phases with FeSe layer distance extended up to ~10.9Å. Further fitting on the resistivity in terms of the Lawrence-Doniach model suggests the superconductivity in the two compounds show obvious quasi-2D characteristics. Intriguingly, a universal sign reverse of Hall coefficient ($R_H$) from negative to positive is observed at ~ 185 K in both superconducting phases with increasing temperatures, regardless of their differences in symmetry and doping levels. First principle calculations revealed the increase in FeSe layer distance will reconstruct the Fermi surface and generate a new hole pocket around Γ point in the Brillouin Zone, which may explain why the missing 'hole pockets' reemerged in these low dimensional FeSe-derived superconductors at high temperatures.

## II. Experimental Method

All sample manipulations were carried out in an argon-filled dry box with an $O_2$ and $H_2O$ content below 1 ppm. Tetragonal $Fe_{1+\delta}Se$ was synthesized following the method described in Ref. [17]. Anhydrous 1,3-Diaminopropane (1,3-DIA) (Sinopharm Chemical Reagent, 99.0% purity) was purified following the method in Ref. [27]. Polycrystalline $Li_x(C_3H_{10}N_2)_yFeSe$ (x = 0.15, 0.2, 0.25, 0.4, 0.6) samples were prepared by solvent-thermal method. In a typical synthesis, 0.0045 mole $Fe_{1+\delta}Se$ powder and ratio of the nominal mole Li pieces (Alfa Aesar, 99.9% purity) and 10 mL ultra-dried 1,3-Diaminopropane were placed in a silica ampoule and then sealed. The ampule was heated at 473.15 K for 12 h in an oven, followed by opening the ampoule in an argon-filled dry box and thoroughly rinsing with fresh 1,3-Diaminopropane. Finally, the solvent was removed under vacuum, and the dark black product was loaded into the measurement cell for structure and physical properties characterization.

X-ray powder diffraction patterns were used for the purpose of phase characterization and structure solution. Room temperature power X-ray diffraction (PXRD) spectra of nominal $Li_x(C_3H_{10}N_2)_yFeSe$ (x = 0.15, 0.2, 0.25, 0.4, 0.6) were collected by using of a PANalytical X'pert Pro diffractometer with Cu K$\alpha$ radiation (40 kV, 40 mA) and a graphite monochromator in a reflection model ($2\theta$ = 5° to 80°, step = 0.017° ($2\theta$)). Well-grounded fine powder samples on glass slides were loaded into a homemade airtight accessories to prevent oxidization of the sample during collecting the diffraction pattern. Laboratory *in situ* PXRD measurements were made using a Rigaku SmartLab instrument (Cu K$\alpha$ radiation) equipped with an Anton Paar HTK-600N Oven Sample stage ($10^{-2}$ Pa, 80 K - 300 K). The room-temperature diffraction pattern (at 300 K) was firstly obtained as a standard, and the low temperature data were collected over the $2\theta$ ranges 5-100° with a step size of 0.017° at 85K, 130K, 150K, 170K, 200K, 250K, respectively. Structure determination and rietveld refinements were performed using Fullprof suites. For solving the structures, $C_3H_{10}N_2$ molecular and FeSe layers were used as independent motifs in a simulated annealing approach, a preliminary structural model is built up with space group P-4 (phase I) and P4/nmm (phase II), and then Li position is located by Fourier difference analysis. Finally, Rietveld refinement against the PXRD data is performed based on this structure model, with the site occupancies constrained to the sample composition during the refinement.

The sample composition (Li: Fe: Se) were determined by inductively coupled plasma mass spectrometry (ICP-AES). The nitrogen contents in the samples were determined using an Oxygen and Nitrogen Analyzer (ONA, Senbao TN-306, Shanghai), with a stand deviation of 0.1 ppm and Nitrogen concentration range between 0.0005% ~ 20%.

Magnetization and resistivity measurements were carried out using a SQUID PPMS-9 system (Quantum Design). Magnetic susceptibility measurements were made in *d.c.* fields of 40 Oe in the temperature range 10-300 K after cooling in zero applied field (ZFC) and in the measuring field (FC). To ensure metallic contact between the polycrystalline grains, the resistivity and Hall measurement were conduct based on presurized samples that are further annealed at 150 °C for 12 hours. Temperature dependence of the resistivity $\rho(T)$ of $Li_x(C_3H_{10}N_2)_{0.32}FeSe$ was measured in a standard four-probe configuration with the applied current less than 2 mA. The Hall coefficient ($R_H$) of $Li_x(C_3H_{10}N_2)_{0.32}FeSe$ were obtained by a linear fit of $\rho_{xy}$ and B from -9 T to 9 T in the temperatures 50 K, 100 K, 150 K, 200 K, 250 K and 300 K, respectively

The first principles calculations were performed with the Vienna ab-initio simulation package (VASP) [28, 29]. We employed the generalized gradient approximation (GGA) in the form of the Perdew-Burke-Ernzerhof (PBE) for the exchange-correlation potentials [30]. The energy cut-off for the plane wave expansion is 500 eV. The Brillouin zones are sampled by Monkhorst-Pack method with meshes of $9\times9\times12$. The structural model of FeSe and $Li_{0.15}(C_3H_{10}N_2)_{0.32}FeSe$ are based on the experimental structure parameters. The model of $(C_3H_{10}N_2)_{0.32}FeSe$ is made by omitting the Li in $Li_{0.15}(C_3H_{10}N_2)_{0.32}FeSe$, which is appropriate to evaluating the influence of expanded layer distance on electronic structure. To simplify calculations, we omitted the charge neutral 1,3-***DIA*** molecules. Doping was simulated by adding extra electrons to the system, together with a compensating uniform positive background. The added electrons is 0.15 *extra* electrons per Fe atom compared with stoichiometric FeSe, in consistent with experimental result.

### III. Results and Discussion
### A. Superconducting Phases

Figure S1(a) shows the PXRD patterns of the as-synthesized compounds co-intercalated by 1,3-*DIA* molecules and Li metal ions with nominal compositions $Li_x(C_3H_{10}N_2)_yFeSe$ (x = 0, 0.15, 0.2, 0.25, 0.4, and 0.6). With the assistance of highly reductive lithium solution (0.15< x < 0.6), a series of new phases with much enhanced lattice parameters appeared. The first phase pure sample without residue FeSe is obtained at x = 0.15 (phase **I**). Based on extinction conditions, the diffraction peaks of phase **I** is indexed into a primitive tetragonal cell, with lattice parameters $a$ = 3.8136(4) Å and $c$ = 10.617(1) Å. With increasing Li concentration (0.15<x<0.25), the initial set of reflections rapidly diminished and were replaced by a new set of reflections. The new phase (phase **II**) without residue phase **I** is firstly obtained at x=0.25, and the pattern was indexed into a primitive tetragonal supercell, with lattice parameters $a$ = 3.7906(2) Å and $c$ = 21.632(1) Å. Strikingly, the diffraction profiles of the new tetragonal phase persist up to x = 0.4, suggesting the phase **II** tolerates a range of Li concentration between x = 0.25 to 0.4. Finally, phase II decomposed at sufficiently high Li concentration (x>0.4), with considerable amount of impurities $Li_2Se$ appeared at x=0.6. The lattice parameters of the phases at different nominal x were summarized in Table I. It is worth noting that the tunable dopant concentration in $Li_x(C_3H_{10}N_2)_yFeSe$ (phase II, 0.25 ≤ x ≤ 0.4) is in contrast to the discreet metal concentrations found in known intercalated FeSe systems [10, 11], suggesting the organic molecule intercalated FeSe may serve an ideal platform to investigate the influence of carrier concentration to superconductivity in FeSe system.

As shown in Table I, the results of ICP-AES analyses indicate that slight lithium loss occurred during the reaction. Meanwhile, the ONA analyses indicated that the intercalated molecules are almost invariant with increasing *x*, assuming that all nitrogen species came from $C_3H_{10}N_2$ molecules. The measured composition of the two phase pure samples at x = 0.15, 0.25~0.4 are Li$_{0.13}$(C$_3$H$_{10}$N$_2$)$_{0.32}$FeSe and Li$_x$(C$_3$H$_{10}$N$_2$)$_{0.33}$FeSe(x=0.22~0.34), respectively.

Table I. Chemical analyses and crystallographic parameters from indexing the PXRD data at 298 K for Li$_x$(C$_3$H$_{10}$N$_2$)$_y$FeSe

| Nominal x | ICP-AES x | y | Symmetry | $a$ (Å) | $c$ (Å) |
|---|---|---|---|---|---|
| 0.15 | 0.13 | 0.32 | *P*-4 | 3.8136(4) | 10.617(1) |
| 0.20 | 0.18 | 0.32 | *Phase I+II* | -- | -- |
| 0.25 | 0.22 | 0.33 | *P* 4/*nmm* | 3.7906(2) | 21.632(1) |
| 0.40 | 0.34 | 0.32 | *P* 4/*nmm* | 3.7858(1) | 21.904(5) |
| 0.60 | 0.42 | 0.32 | *Phase II+Li$_2$Se* | 3.7878(3) | 21.912(4) |

**B. Crystal structures**

As shown in Fig. 1(a), the crystal structure of Li$_{0.15}$(C$_3$H$_{10}$N$_2$)$_{0.32}$FeSe (phase I, space group *P*-4) was resolved by *ab initio* structure determination from PXRD data. Rietveld refinement is performed based on this structure model and produced a satisfying fit to the diffraction pattern, with R$_p$ = 2.35% and R$_{wp}$ = 3.56% at 295 K. As shown in Figure 1a, the intercalated C$_3$H$_{10}$N$_2$ molecules were centered on general crystallographic positions 4h site: (x, y, z), and disordered over four diagonal orientations. Under the refined structure model, the two shortest H-Se distances in the compound are 2.765(4) Å and 2.779(2) Å, respectively, in consistence with normal hydrogen bonding interactions [10, 31]. The two Fe atoms are located at 1a site: (0, 0, 0) and 1c site: (1/2, 1/2, 0), and the Se atoms are located at 2g (0, 1/2, z) site. Li ions were located adjacent to the FeSe layers at sites 2g (0, 1/2, z) site. Moreover, the compressed FeSe$_4$ tetrahedra found in Fe$_{1+\delta}$Se are retained, with Se-Fe-Se bond angles of 104.0(1)° and 112.3(1)° compared with values of 103.9(2)° and 112.3(4)° in Fe$_{1+\delta}$Se [17].

As shown in Figure 1(b), the crystal structure of Li$_{0.25}$(C$_3$H$_{10}$N$_2$)$_{0.33}$FeSe (phase II) is successfully determined in a similar manner in space group *P*4/*nmm*. The main structural difference between the two phases lies in that with increasing Li concentration, the orientations of the intercalated molecules in phase II become inequivalent between the adjacent FeSe layers. Subsequently, the unit cell is doubled along *c* axis. Along with the phase transition, the FeSe interlayer distance is further extended from 10.617 Å to 10.816 Å. There is also hydrogen bonding between the hydrogen atoms and the anion of FeSe layers, with the short H-Se bond distances around 2.63 Å ~ 2.91 Å [10, 31]. Moreover, the FeSe$_4$ tetrahedra in phase II become more distorted with increasing electron doping. As shown in Figure 1, the Se-Fe-Se bond angles increased from 112.31° to 112.51°, the values are comparable to the results in ammonia intercalated Li$_{0.6(1)}$(ND$_2$)$_{0.2(1)}$(ND$_3$)$_{0.8(1)}$Fe$_2$Se$_2$ (112.84(3)°) [10]. As shown in Figure 1e, we also refined the crystal structure of phase II compound at x=0.4, with the final reliability factors reached R$_p$ = 3.89% and R$_{wp}$ = 5.16%. The refined structural parameters confirmed that the doping continuously increased the distortion of FeSe$_4$ tetrahedra, with the Se-Fe-Se angles further increased to 112.61°. Meanwhile, the anion height within the FeSe layers also monoclinic increased from 1.491(6) Å (*x*=0.15) to 1.497(3) Å (*x*=0.4).

### C. Low-temperature crystal structures

For iron pnictides, a tetragonal-to-orthorhombic structural transition (from *C*4 symmetry to *C*2 symmetry) is often observed in the under-doped states. To determine the evolution of Li$_x$(C$_2$H$_8$N$_2$)$_{0.5}$Fe$_2$Se$_2$ crystal structure at low temperatures, X-ray powder-diffraction data were collected for the least electron doped sample (x=0.15) at T = 85 K, 110 K, 130 K, 150K, 170K, 200K and 295 K, respectively. Figure S2~8 shows the observed, calculated, and difference diffraction profiles for Li$_{0.15}$(C$_3$H$_0$N$_2$)$_{0.32}$Fe$_2$Se$_2$. The low temperature patterns can all be well fitted by the room temperature structure model in the space group *P*-4, indicating no 'nematic' phase transition occurred between 85 K and 250 K. The temperature dependence of unit-cell parameters *a* and *c* values, the Se-Fe-Se bond angles, Fe-Se bond lengths as well as the anion height at the measured temperatures are plotted in Figure S9. From 85 K to room temperature, the lattice parameter *a* increases only by 0.075%, whereas the lattice parameter *c* expand more obvious by 0.25%. The temperature dependent structural details are present in Figure S9.

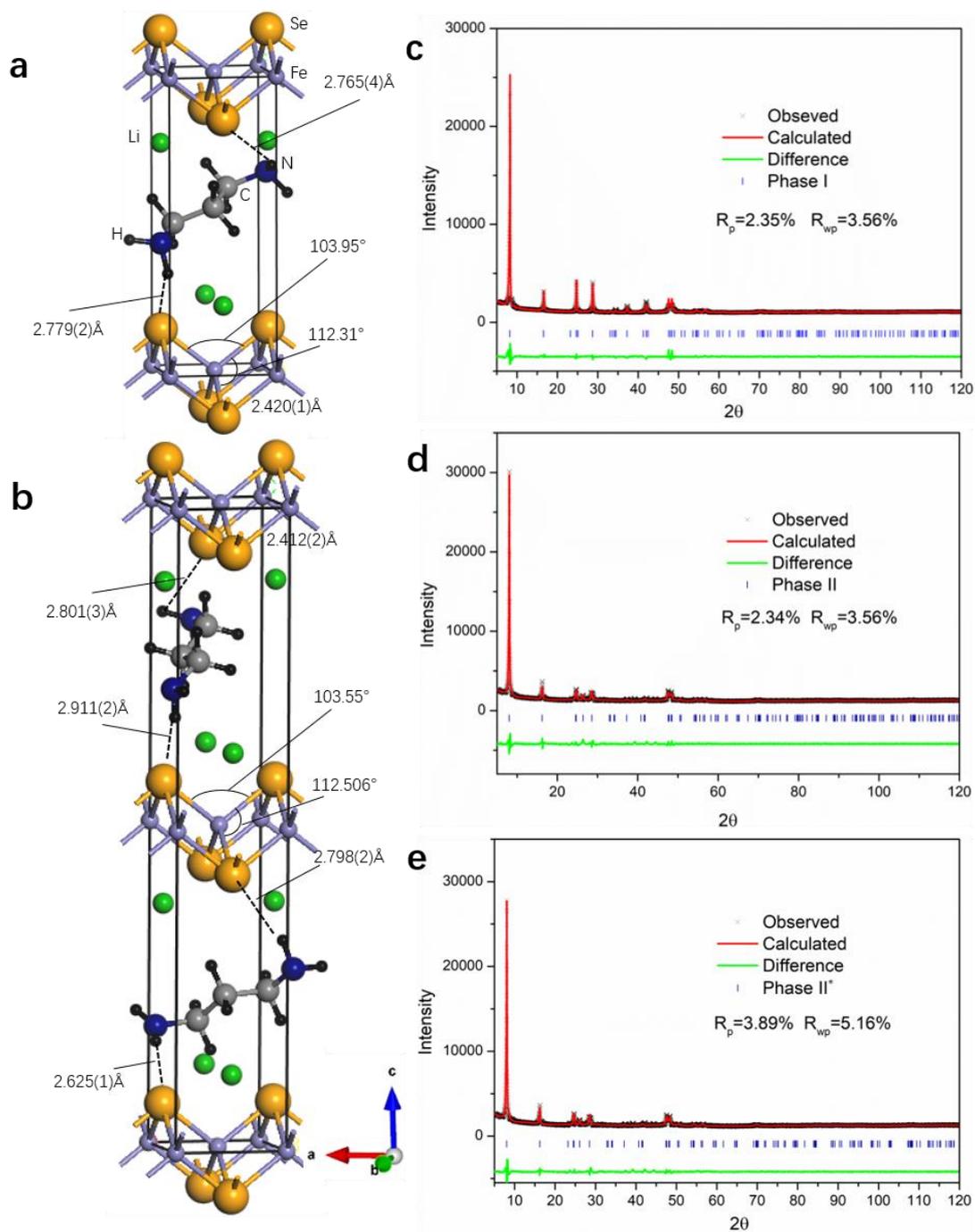

Figure 1. Crystal structures and Rietveld refinement for two phases in 1,3-DIA intercalated FeSe at 300 K. (a) Structural model of phase I. (b) Structural model with orientational disorder of the molecules for phase II ($x$=0.25). (c) Refinement against powder X-ray diffraction data of phase I ($x$=0.15). (d) Refinement against powder X-ray diffraction data of phase II ($x$=0.25). (e) Refinement against powder X-ray diffraction data of phase II ($x$=0.4).

## D. Quasi-2D Superconductivity behaivior

Figure 2(a) shows the temperature dependence of zero-field cooling (ZFC) magnetic susceptibility of three phase pure $Li_x(C_2H_8N_2)_yFe_2Se_2$ ($x$ = 0.15, 0.25, 0.4) samples under an external magnetic field of 10 Oe. With the carrier doping from intercalated Li, sharp transitions from normal state to superconducting state are observed at temperature range from 40 K ~ 46 K in the measured samples. The as-synthesized $Li_{0.15}(C_3N_2H_{10})_{0.33}Fe_2Se_2$ shows a diamagnetic transition at $T_c$ up to 40 K with a considerable shielding fraction of 49% at 10 K in the zero-field-cooling condition. Further confirmation of superconductivity is shown in Figure 2b, which shows the temperature-dependent electrical resistivity measured on cold-pressed pellets. Apart from the metallic normal state above the $T_c$ in $Li_{0.15}(C_3N_2H_{10})_{0.32}Fe_2Se_2$, a rapid decrease of resistivity appeared around 40 K, and zero resistivity was reached at 36 K. With further increased doping concentration in phase II, the maximum $T_c$ up to 46 K is realized in $Li_{0.25}(C_3N_2H_{10})_{0.33}Fe_2Se_2$, equal to the optimal $T_c$ value of the ammonia-intercalated FeSe. Correspondingly, the electrical resistivity shows transition around 45K and zero resistivity is realized at 41 K. From $x$ = 0.25 to 0.4, the $T_c$ is found slightly depressed to 44 K at high Li concentration, indicating the system entered into an over-doped region. Meanwhile, the superconducting volume fraction continually increased with x for two Phase **II** samples, with their maximum value reached 76%, confirming the bulk superconductivity. Our results clearly show that the superconductivity in tetragonal $Li_x(C_2H_8N_2)_yFe_2Se_2$ is correlated to the carrier concentration.

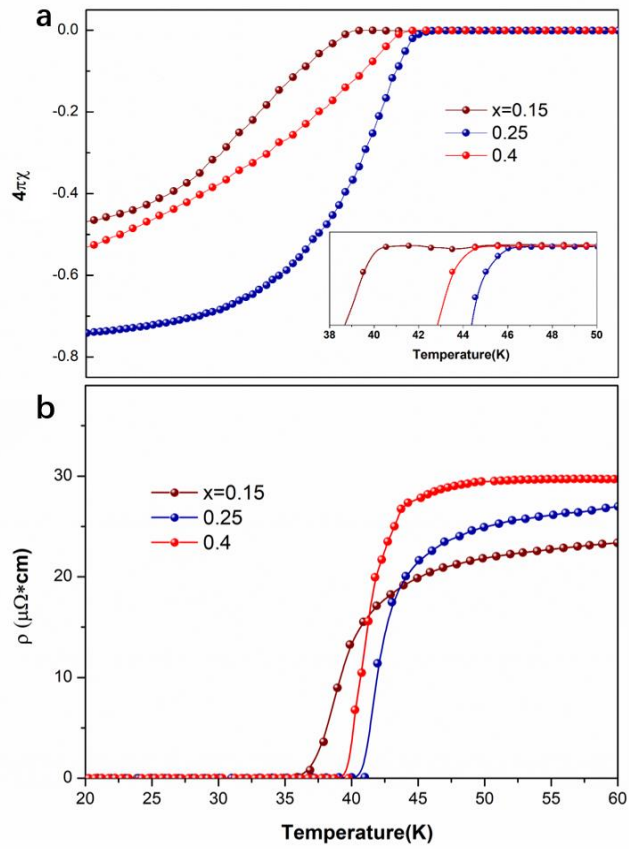

Figure 2. (a) Magnetic susceptibilities of three phase pure Li$_x$(C$_3$H$_{10}$N$_2$)$_y$FeSe ($x$ = 0.15, 0.25, 0.4) and the inset is the enlarged view. (b) Electrical resistivity of three phase pure Li$_x$(C$_3$H$_{10}$N$_2$)$_y$FeSe ($x$ = 0.15, 0.25, 0.4) samples.

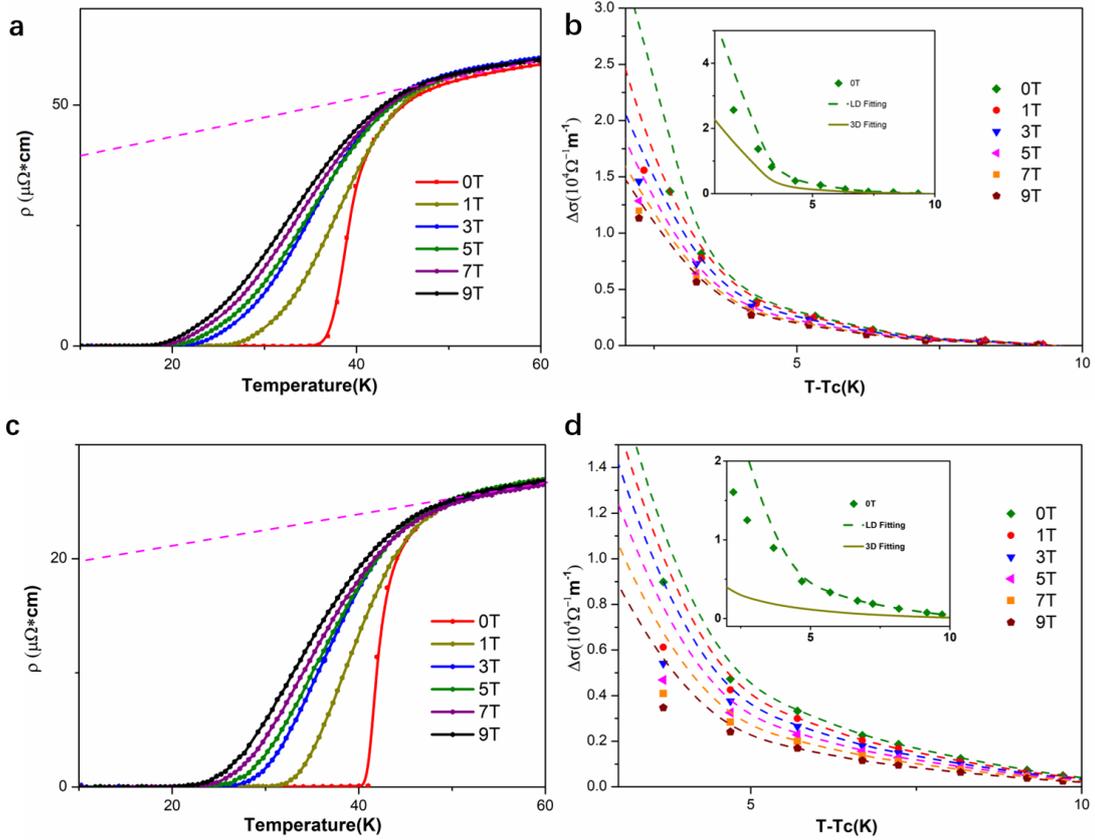

Figure 3. Quasi-2D superconductivity behaivior of $Li_x(C_3H_{10}N_2)_{0.32}FeSe$ (x=0.15, 0.25). (a) & (c) Temperature dependence of the resistivity around Tc of $Li_x(C_3H_{10}N_2)_{0.32}FeSe$ (x=0.15, 0.25) under increasing magnetic fields. The lines are examples (for $\mu_0H$ = 9 T) of the background contribution, as determined by a linear fit above 42 K, where fluctuation effects are expected to be negligible. (b) & (d) The fluctuation contribution to the conductivity $\Delta\sigma$ varies with temperature under external magnetic fields up to 9T for $Li_x(C_3H_{10}N_2)_{0.32}FeSe$ (x=0.15, 0.25). The inset is fitting of $\Delta\sigma$ vs temperature with the quasi 2D Lawrence-Doniach model (green dash line) and 3D anisotropic Ginsburg-Landau model (brown solid line).

For the $Li_x(C_3H_{10}N_2)_{0.32}FeSe$ samples, the intercalation of 1,3-DIA molecule significantly enlarged the distance between the adjacent superconducting FeSe layers, which will weaken the correlation strength along *c*-axis for this FeSCs system. Figure 3 displays the field dependent electric resistivity of $Li_x(C_3H_{10}N_2)_{0.32}FeSe$ (x=0.15, 0.25). As a signature of the quasi two-dimensional (2D) superconductivity in $Li_x(C_3H_{10}N_2)_{0.32}FeSe$, the resistivity transition near $T_c^{onset}$ becomes too smooth to define a mean-field transition point, especially in the presence of a magnetic field. The fluctuation magnetoconductivity above the superconducting transition temperature (Tc) is hence analyzed in terms of the Lawrence-Doniach (LD)(equation (1) & (2)) approach for 2D superconductivity [32, 33].

$$\Delta\sigma_{LD} = \frac{e^2}{64\pi\hbar h}\frac{1}{s}\int_{-\pi/s}^{\pi/s}dk_z\left[\Psi^1\left(\frac{\varepsilon+h+\omega_{k_z}^{LD}}{2h}\right)-\Psi^1\left(\frac{c+h+\omega_{k_z}^{LD}}{2h}\right)\right] \quad (1)$$

$$\omega_{k_z}^{LD} = \gamma\left[1-cos(k_zs)\right]\Big/2 \quad (2)$$

Here $\psi^1$ is the first derivative of digamma function, e is electron charge, $h$ is the reduced magnetic field, $\varepsilon$ is $\ln(T/T_c^{mid})$, $s$ is the distance between adjacent FeSe layers and $r$ is a fitting parameter corresponding to coherence length amplitude $\zeta(0)$. Figure 3b presents the equation contribution to the conductivity $\Delta\sigma$ which varies with temperature under external magnetic fields from 0 T to 9 T. The fittings lead to an averaged $\zeta(0)$ = 0.9652Å, a value one order of magnitude smaller than the FeSe inter-layer distance. As shown in the inset of Fig. 3(b), the experimental $\Delta\sigma$ strongly deviates from the 3D anisotropic Ginsburg-Landau (GL) model, and agrees well with the Lawrence-Doniach model, clearly showing the superconductivity of $Li_{0.15}(C_3H_{10}N_2)_{0.32}FeSe$ is quasi-2D in nature. Moreover, the fitting also yields a zero-temperature upper critical field $H_{c2}(0)$ = 89T. As for $Li_{0.25}(C_3H_{10}N_2)_{0.33}FeSe$, a similar quasi-2D superconducting behaivior also clearly existed. The fitting based on Lawrence-Doniach function yield a coherence length amplitude $\zeta(0)$=0.8425 Å, and the zero- temperature upper critical field $H_{c2}(0)$ is obtained as 82 T.

### E. Hall resistivity

In Figure 4, we show the transport properties of $Li_x(C_3H_0N_2)_yFe_2Se_2$ ($x$ = 0.15, 0.25, 0.4) samples at normal state. In the whole temperature region, Hall resistivity $\rho_{xy}(\mu_0H)$ of $Li_{0.15}(C_3H_{10}N_2)_{0.32}FeSe$ samples show good linear relation against magnetic field up to 9 T. The derived Hall coefficients $R_H = \rho_{xy}/\mu_0H$ at 9 T exhibits strong temperature dependent (Figure 4a). The $R_H$ is negative below 200 K and the absolute values decrease rapidly with increasing temperature. Finally, the $R_H$ becomes positive at higher temperature, i.e., there is a sign change ~ 185 K. It suggests the existence of two different types of charge carriers in $Li_{0.15}(C_3H_{10}N_2)_{0.32}FeSe$, the dominant carriers are electron-type at the low temperatures and become hole-type at high temperature. For instance, the apparent carrier concentration $n_H$ (=$1/eR_H$ = $n_h$−$n_e$, where $n_h$ and $n_e$ are the carrier concentrations of hole and electron pockets) can be obtained as $-1.48 \times 10^{21}$ cm$^{-3}$ at 50 K and $2.1 \times 10^{21}$ cm$^{-3}$ at 300 K for $Li_{0.15}(C_3H_{10}N_2)_{0.32}FeSe$.

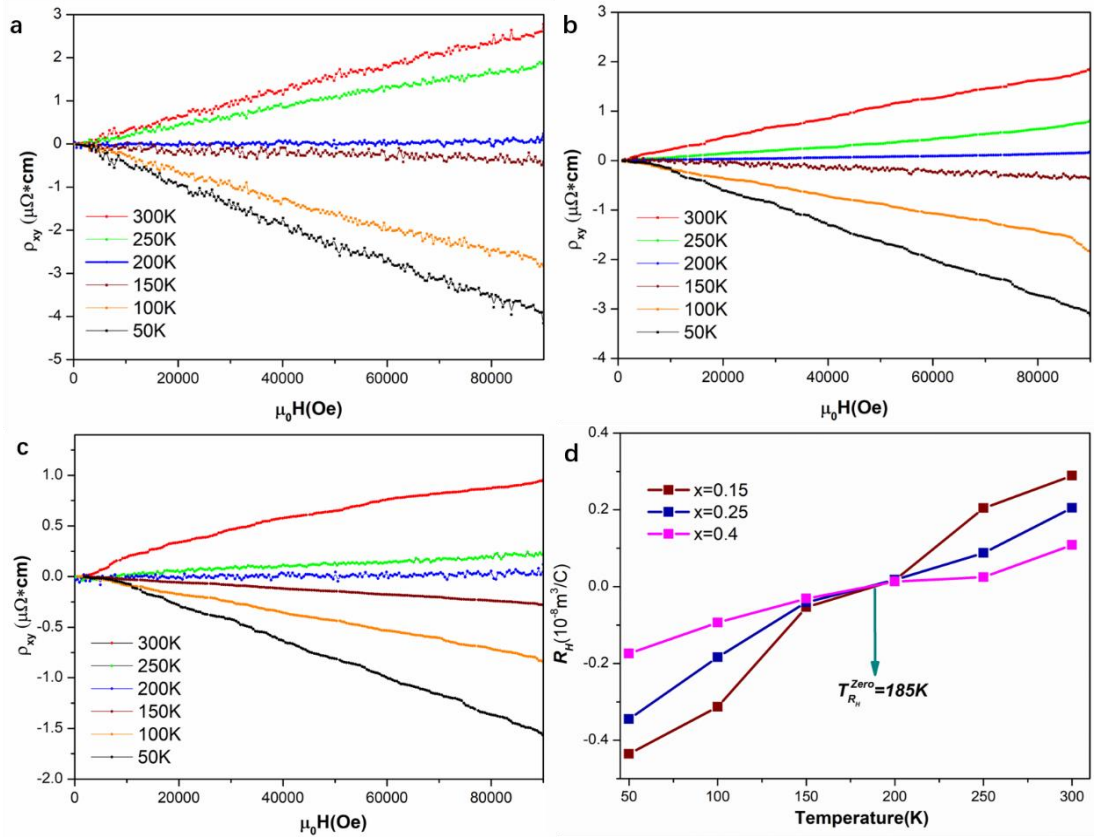

Figure 4. Hall effect transverse resistivity $\rho_{xy}$ measurement. (a), (b) and (c) is the field dependence of Hall effect transverse resistivity—$\rho_{xy}$ for $Li_x(C_3H_0N_2)_yFe_2Se_2$ (x = 0.15, 0.25, 0.4) samples. (d) Temperature dependence of the Hall coefficient for three samples, $R_H$ changes sign from negative to positive at a similar point about 185K, demonstrating dominant hole carriers appeared above 185 K.

Figure 4(b) and (c) show the Hall resistivity $\rho_{xy}(\mu_0H)$ of two phase II samples with much higher dopant concentration (x = 0.25 and 0.4). The Hall resistivity $\rho_{xy}(\mu_0H)$ of the two samples show similar linear dependance on magnetic field. The increased Li concentration significaltly decreased the absolute value of $\rho_{xy}(\mu_0H)$ of the samples above $T_c$. It indicates that the apparent carrier concentration $n_H$ in optimum doped

Li$_{0.25}$(C$_3$H$_{10}$N$_2$)$_{0.33}$FeSe (-1.82 × 10$^{21}$ cm$^{-3}$ at 50 K) and overdoped Li$_{0.40}$(C$_3$H$_{10}$N$_2$)$_{0.32}$ FeSe (-3.3 × 10$^{21}$ cm$^{-3}$ at 50 K) is much larger than the phase I sample, in consistant with the increased Li concentration in the two compounds. More strikingly, as shown in Figure 5(d) and Table II, the $R_H$ of both the samples also increase rapidly with elevating temperatures, and changed their sign at almost the same temperature around 185 K. The universial sign change of $R_H$ in Li$_x$(C$_3$H$_{10}$N$_2$)$_y$FeSe ($x$ = 0.15, 0.25, 0.4) samples suggest the reconstruction of fermi surface topology at high temperatures are a common phenomenon in propylenediamine intercalated FeSe system.

It has to be mentioned that for electron-doped FeSe, the sign change of R$_H$ at high tempratures is very uncommon. Hall coefficient of both K$_x$Fe$_{2-y}$Se$_2$ [22] and (Li$_{0.8}$Fe$_{0.2}$)OHFeSe single crystals [23, 24] are negative and exhibit concave shape of $R_H$ in the measured tempretaure range. Meanwhile, the $R_H$ of superocnducting Li$_x$(NH$_3$)$_y$FeSe are also negative below RT but increased monotonically with tempature [34]. The negative $R_H$ in heavy electron-doped FeSe is consistent with their novel electronic structure consist of only electron pockets [5]. Only in Li-NH$_3$ intercalated Li$_x$(NH$_3$)$_y$FeTe$_{0.6}$Se$_{0.4}$ [35], Li *et al.* recently reported a sign reversal of $R_H$ at high temperature. However, this results is not suprizing considering the dominate carriers in parent compound FeTe$_{0.6}$Se$_{0.4}$ is hole-type. Furthermore, the apparent carrier concentration $n_H$ in Li$_x$(NH$_3$)$_y$FeTe$_{0.6}$Se$_{0.4}$ is significant smaller (-2.8×10$^{20}$ cm$^{-3}$ at 30 K) than its FeSe analogue Li$_x$(NH$_3$)$_y$FeSe (-1.3×10$^{21}$ cm$^{-3}$ at 50 K) [34,35], it is hence proposed that in intercalated FeTe$_{0.6}$Se$_{0.4}$, the hole bands may be still crossing the Fermi energy level (E$_F$) at low temperatures. Recently, a $^{77}$Se, $^7$Li, and $^1$H nuclear magnetic resonance (NMR) study of the ethylenediamine intercalated Li$_x$(C$_2$H$_8$N$_2$)$_y$ Fe$_{2-z}$Se$_2$ identified strong temperature dependence of $^{77}$Se NMR shift and spin-lattice relaxation rate, 1/$^{77}T_1$ [36]. The phenomenon was attributed to the hole like bands moving close to the Fermi energy at high temperatures. Here, the significant sign reversal of R$_H$ universially observed in the title superconductors proved more direct evidences that at least one of the hole pockets beneath the E$_F$ can crossed the E$_F$ at high temperuates, and leads to a fundamental change in the fermi surface topologies.

Table II. The Hall coefficients from 50K to 300K for x=0.15, 0.25 and 0.4

| T (K) | x=0.15 | | x=0.25 | | x=0.4 | |
|---|---|---|---|---|---|---|
| | $R_H$ ($10^{-3}$ cm$^3$/C) | $n_H$ ($10^{21}$ cm$^{-3}$) | $R_H$ ($10^{-3}$ cm$^3$/C) | $n_H$ ($10^{21}$ cm$^{-3}$) | $R_H$ ($10^{-3}$ cm$^3$/C) | $n_H$ ($10^{21}$ cm$^{-3}$) |
| 50 | -43.55 | -1.48 | -34.51 | -1.81 | -17.41 | -3.59 |
| 100 | -31.33 | -1.99 | -18.35 | -3.41 | -9.35 | -6.68 |
| 150 | -5.34 | -11.82 | -4.06 | -15.37 | -3.12 | -20.08 |
| 200 | 1.66 | 37.51 | 1.77 | 35.15 | 1.33 | 46.88 |
| 250 | 20.47 | 3.05 | 8.75 | 7.14 | 2.46 | 25.34 |
| 300 | 28.89 | 2.16 | 20.51 | 3.05 | 10.82 | 5.78 |

## F. Electronic structure

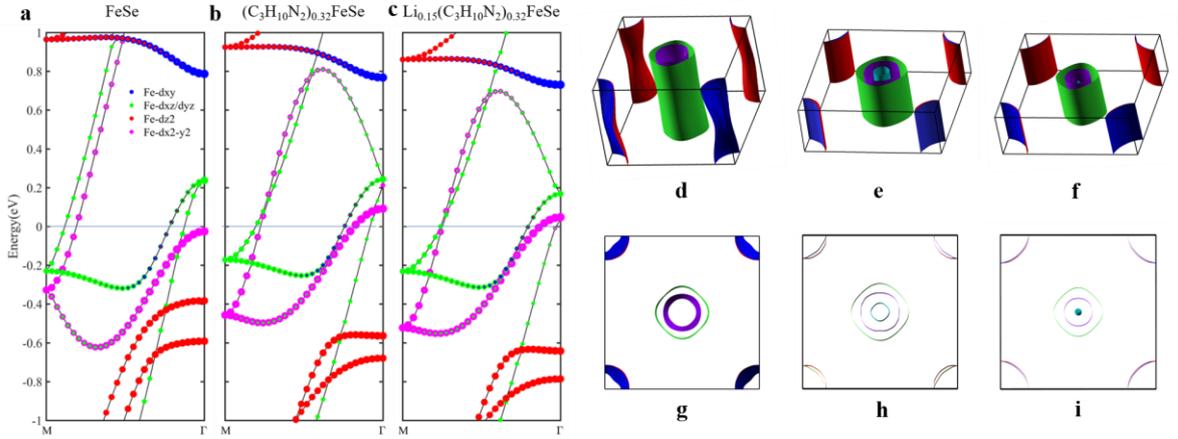

Figure 5 Electronic structure of bulk FeSe, $(C_3H_{10}N_2)_{0.32}$FeSe and Li$_{0.15}$(C$_3$H$_{10}$N$_2$)$_{0.32}$FeSe. (a~c) Band structure along M to Γ route of bulk FeSe, (C$_3$H$_{10}$N$_2$)$_{0.32}$FeSe and Li$_{0.15}$(C$_3$H$_{10}$N$_2$)$_{0.32}$FeSe. (d~f) Fermi surface of bulk FeSe, (C$_3$H$_{10}$N$_2$)$_{0.32}$FeSe and Li$_{0.15}$(C$_3$H$_{10}$N$_2$)$_{0.32}$FeSe (Γ point is at the center and M point is at the corner). (g~i) Vertical view of the Fermi surface. The Fermi energy is set to zero.

To clarify the influence of molecule intercalation and electron doping on the electronic structures, band structures of FeSe, Li$_{0.15}$(C$_3$H$_{10}$N$_2$)$_{0.32}$FeSe and (C$_3$H$_{10}$N$_2$)$_{0.32}$FeSe (charge neutral) were calculated by first principles calculations. The orbital resolved DFT band structures along M to Γ route are presented in Figure 5, and the more complete band structures are shown in Figure S17. For bulk FeSe, there are four bands crossed the Fermi level, with two hole-like bands around the zone center (M point) and two electron-like bands around the zone corner (Γ point). Figure 5 (d) & (g) show the Fermi surface for bulk FeSe, the volumes enclosed by the Fermi surface are found to be 0.230 holes/cell and 0.231 electrons/cell, respectively. The results corresponding to a carrier density of $2.93\times10^{21}$ holes/cm$^3$ and $2.96\times10^{21}$ electrons/cm$^3$, in consistence with the almost compensate electronic and hole carriers in bulk FeSe. The results are similar to the previous calculations and direct ARPES measurements results [37,38].

By expanding the FeSe layer distance up to 10.62Å, the band structure of $(C_3H_{10}N_2)_{0.32}$FeSe exhibits a nearly flat dispersion along F-Z route with quasi-2D character, suggestting the dispersion along *c*-axis is weakened as the molecule intercalated (Figure S17). Meanwhile, the intercaltion has a profound influence on the electronic structure around Γ point. As shown in Figure 5, bulk FeSe has two hole pockets at Γ point, which are dominated by $d_{xz/yz}$ orbital. Notably, the molecule intercaltion in $(C_3H_{10}N_2)_{0.32}$FeSe trigger a new $d_{x2-y2}$ hole pocket above the fermi level around Γ point (Figure 5(b)). Apart from this, an empty-state band domainated by $d_{xz/yz}$ orbital sinked significantly as the FeSe layers are expanded. This sinked empty band was recently observed through STM in another low-dimentinoal FeSe system (1-layer FeSe film on SrTiO$_3$ substrate) [39]. Figure 5 (e) & (h) shows the Fermi surface for $(C_3H_{10}N_2)_{0.32}$FeSe, the volume enclosed by the hole-like Fermi surface are found to increased significantly compared to bulk FeSe, meanwhile, the electronic Fermi surface is almost unchanged. In particular, the volumes enclosed by Fermi surface are 0.791 holes/cell and 0.525 electrons/cell, namely the carrier concentration are $5.12\times10^{21}$ holes/cm$^3$ and $3.41\times10^{21}$ electrons/cm$^3$, respectively. The results suggest that the quasi 2D structure will reconstruct the Fermi surface around Γ point and significantly increase the hole carrier concentration.

Futhure calculation of $Li_{0.15}(C_3H_{10}N_2)_{0.32}$FeSe shows that the three hole pockets still existed around Γ, but compared to $(C_3H_{10}N_2)_{0.32}$FeSe their volume are suppressed by electron doping. Meanwhile, the new Γ-centered empty-state band continue to sink with increased electronic doping level. In fact, under much heavier electron doping this empty-state band will across the fermi level and cause a lifshitz transition, as observed in a recent ARPES measurements [16]. Figure 5 (f) & (i) shows the Fermi surface for $Li_{0.15}(C_3H_{10}N_2)_{0.32}FeSe$, the volumes enclosed by Fermi surface are 0.341 holes/cell and 0.682 electrons/cell, which corresponding to carrier concentration of $2.214\times10^{21}$ holes/cm$^3$ and $4.421\times10^{21}$ electrons/cm$^3$, respectively. The result is in line with Hall measurements shown in Table II, wherein the domantiate carrier in $Li_{0.15}(C_3H_{10}N_2)_{0.32}FeSe$ is electronic at low temperature. But as the temperature rose, the chemical potential decreased and the Fermi level continually sinks. The already enlarged hole pockets due to expanded FeSe layer around Γ point will get further enhanced. That explained the trend of increased hole carrier density in various low dimentional FeSe systems, such as LiNH$_3$FeSe and LiNH$_3$FeSeTe[34, 35]. Therefore, the enhanced hole pockets in low dimensional FeSe systems and the decreased chemical potential with elevating temperature are proposed to be responsible for the sign reversal of hall resistivity in $Li_x(C_3H_0N_2)_yFe_2Se_2$ (x = 0.15, 0.25, 0.4).

It has been noted that a universal sign reversal of Hall resistance with temperature is previously observed in FeSe films on SiTO$_3$ substrate, the 1-2 unit cell (UC) thick films demonstrating a similar crossover from hole conduction to electron conduction above $T_c$ [26]. More intriguing, the sign reversal temperature is found directly correlated to the layer thickness, suggesting the sign reversal of Hall resistance could be related to the decrese of dimensionallity in the FeSe system. Indeed, as shown in Table I, the interlayer distances of large organic molecule intercalated FeSe are significantly larger than other known FeSe based superconductors, and consequently the Fermi surface of those materials should more close to the 2D limits as in one unit cell FeSe films. Considering the similar band structures in intercalated FeSe and monolayer FeSe, it is postulated that the same picture may also be valid to explain the observation of even higher $T_c$ up to 63 K in highly strained 1 ML FeSe on the rectangular (100) face of rutile TiO$_2$[40]. Moreover, it is expected the sign reversal of Hall resistance should be universal for a majority of organic molecule intercalated FeSe materials with super-expanded FeSe interlayer distances.

## IV. CONCLUSION

In summary, two tetragonal superconducting phases are identified in $Li_x(C_3H_{10}N_2)_{0.32}FeSe$ (x = 0.15~0.4) system, i.e., $Li_{0.15}(C_3H_{10}N_2)_{0.32}FeSe$ (phase **I**) and $Li_x(C_3H_{10}N_2)_{0.32}FeSe$ (phase **II**, 0.25 < x < 0.4). The lithium concentration is revealed as a key parameter controlling both the crystal structures and the Tc. The determined structures demonstrate that between the FeSe layers, the orientation of 1,3-DIA molecules is coupled with Li concentration, and the FeSe layer distance is expanded up to ~10.9Å. Superconductivity is observed at 40 K in phase **I**, and the highest $T_c$ up to 46K is realized in phase **II** at x =0.25. The Hall coefficient $R_H$ of the two phases are negative at low temperature, indicating dominant electron-type carriers due to Li doping. Meanwhile, a universal sign reversal of $R_H$ at 185K is observed in both phases and at three different doping levels, implying that new hole pockets appears in all the superconducting compounds at high temperature. Fitting on resistivity in terms of the Lawrence-Doniach model suggests the two superconductors belong to the quasi-two dimensional (2D) system. Further DFT calculations revealed that the increase in FeSe layer distance will reconstruct the Fermi surface and generate a new hole pocket around Γ point in the Brillouin Zone.

**Acknowledgments**

This work is financially supported by the National Natural Science Foundation of China under granting numbers: No. 51472266, 51532010, No. 91422303, 51772323; the National Key Research and Development of China (2016YFA0300301); and the Key Research Program of Frontier Sciences, CAS, Grant No. QYZDJ-SSW-SLH013.

Corresponding authors: shifengjin@iphy.ac.cn; l.gu@iphy.ac.cn; chenx29@iphy.ac.cn